\documentstyle[12pt]{article}
\begin{document}
\begin{flushright}
IITM-TH-94-04\\
\vspace{0.2cm}
December 1994~~
\end{flushright}
\vspace{1cm}
\baselineskip=24pt
\begin{center}
{\Large \bf
The Weinberg - Faddeev solution to the \\
problems of Quantum Field Theory and\\
Quantum Gravity:Quantum Spacetime}\\
\vspace{1cm}
{\bf G.H.Gadiyar,\\
Department of Mathematics, Indian Institute of Technology, Madras
600 036, INDIA.}
\end{center}
\vspace{1.5cm}
\noindent {\bf Abstract:} In this paper a fundamental length is
introduced into physics. This is done in a way which respects special
relativity and quantum field theory. The theory has formal similarity to
quantum field theory though its properties are far better: divergences
are got rid of. The problem of quantizing gravity is straightforward in
the approach.

\newpage

Trying to unify the General Theory of Relativity with quantum field
theory has led to great difficulties. In this paper a new and simple
approach to the problem of quantum gravity is sketched.

This work is inspired by two streams of thought: one due to Weinberg [2]
which is rich in physical ideas and the other due to Faddeev [1] which
has deep mathematical thoughts. The merging of these two approaches
yields a simple solution to two theoretical problems: the problem of
divergences in quantum field theory and the issues of quantum gravity.

Weinberg [2] follows an unorthodox approach to the General Theory of
Relativity.  Traditionally General Relativity is based  on the
equivalence principle and the geometry of curved spacetime. Thus the
approach is heavily based on geometry.  According to Weinberg this
approach drives a wedge between General Relativity and particle physics.
He takes a different point of view in two papers written in the
mid-sixties. He bases his arguments on the axioms of quantum field
theory.  He analyses first the case of massless spin one particles and
derives the Maxwell equations. Next he analyses the case of massless
spin two particles.  Surprisingly he is able to derive the equivalence
principle and the Einstein equations from the properties of Lorentz
invariance and quantum field theory. However the problems of
renormalization remain. Weinberg's analysis seems to indicate that we
are victims of history. If quantum field theory had been developed
before Einstein's equations were discovered curved spacetime would not
dominate the thoughts of theoretical physicists. If one can address the
problem of renormalization then gravity can be considered the Lorentz
invariant theory of massless spin two particles coupled to the energy
momentum tensor.

Faddeev [1] recently made an interesting analysis of relativity and
quantum mechanics.  He says that both these developments were
mathematically what are called deformations.  This means that the
dimensionful parameters $c$ the velocity of light and $\hbar $ the
Planck's constant were introduced into the older structures of physics
in a consistent way. He then says that there is one more constant which
is dimensionful, namely, the Planck length $L$. He forcefully argues
that the general structure of physics should be modified to include this
constant as well. As $L^2~=~\displaystyle{\frac{G\hbar}{c^3}}$
where $G$ is the Newton's gravitation constant, one suspects that the
structure of spacetime will need modification to address the problems of
quantum gravity.

Thus the logic of the approach of this paper is as follows: first modify
the structure of quantum field theory (which respects special relativity
and quantum physics) in a consistent way so as to introduce a
fundamental length. To do this requires
heavy use of the correspondence principle: we guess that the new
structure  will be formally similar to the older structure. This is like
the fact the both classical and quantum mechanics have underlying Lie
algebra structures, one the Poisson bracket and the other the
commutator. To quantize gravity the approach pioneered by Weinberg
becomes feasible as it turns out that introducing $L$ cures the
divergence difficulties. Thus the solution of one theoretical problem
can be used to resolve another.

To introduce a fundamental length in a consistent way requires a couple
of assumptions. The definition of interval is modified to
$$
\hat s^2~=~\hat x^{\dagger}_\mu \, \eta _{\mu \nu} \, \hat x_\nu
\eqno(1)
$$
with
$$
[\hat x_\mu \, , \hat x^{\dagger}_\nu] ~=~ L^2 \, \eta_{\mu \nu} \ .
\eqno(2)
$$
Here $x_\mu$ and $x^{\dagger}_\mu$ are annihilation and creation
operators. Thus the interval is quantized in units of $L^2$. However
one has to choose the state on which these operators act. Here we argue
from the correspondence principle that it is possibly a coherent state
\begin{eqnarray*}
\hat x_\mu \, | \, x_\mu > &=& x_\mu \, | \, x_\mu > \ , \hspace{8cm}
(3) \\
\\
<x_\mu \, |\, x_\mu ' > &=& e^{-{\displaystyle{\frac{(x_\mu ~-~ x_\mu')
^2}{2L^2}}}} \, . \hspace{7cm} (4)
\end{eqnarray*}
We consider only real coherent states which have only real $x_\mu$. This
is possible as the set can be used to construct a partition of unit and
still constitutes an overcomplete basis.

In one dimension,
\begin{eqnarray*}
\hat x \, | \, 0> &=& 0 \\
\hat x \, | \, n> &=& \sqrt{n} \, | \, n-1 > \\
\hat x^{\dagger} \, | \, n> &=& \sqrt{n+1} \, | \, n+1> \\
\hat x \, | \, x > &=& x \, | \, x> \\
| \, x> &=& e^{-\displaystyle{\frac{x^2}{2L^2}}} \sum^{\infty}_0
\frac{x^n}{\sqrt{n!}} \, | \, n> \hspace{6.6cm} (5)
\end{eqnarray*}
where $| \, n>$ has eigen value $nL^2$. Further we demand
$$
\int dx \, | \, x><x \, | ~=~ \sum_n  \, | \, n><n \, | ~=~ 1 \, .
\eqno(6)
$$
But for notational changes this is the same as the harmonic oscillator.

To pinpoint what is the effect of this modification let us consider a
scalar field.  Mathematically one has to replace $x_\mu$ in the usual
approach by $ \hat x_\mu \, | \, x_\mu > ~=~ x_\mu \, | \, x_\mu > $ and
hence $\phi(x_\mu)$ by
$\phi (\hat x_\mu) \, | \, x_\mu > ~=~ \phi (x_\mu) \, | \, x_\mu >$ and
$\displaystyle{ \frac{\partial \phi(x_\mu)}{\partial x^\mu}}$ by
$\displaystyle{\frac{\partial \phi(x_\mu)}{\partial x_\mu} \, | \, x_\mu
>}$.
Thus the action is now
$$
S~=~ \int <x \, | \, \partial _\mu \phi(x) \, \partial ^\mu \phi (x) \, | \, x
> dx \ .\eqno(7)
$$
Varying $\phi(x)$ then yields the equation of motion
$$
\delta S ~=~ \int <x \, | \, \delta \phi(x) \, \Box \, \phi(x) \, | \, x
> dx ~=~ 0 \ .\eqno(8)
$$
Now $<x \, | \, \delta \phi (x)$ is an arbitrary vector in the Hilbert
space. Hence if $\delta S$ is zero it follows that
$$
\Box \, \phi (x) \, | \, x > ~=~ 0 \ . \eqno(9)
$$
This is akin to the Schr\"{o}dinger equation. This equation replaces the
usual $ \Box \phi (x) ~=~0 $. Further  one has the analog of $e^{ikx}$
which in Dirac's notation is $<k \, | \, x>$. Here it can be seen that
\begin{eqnarray*}
\int dk <x \, | \, k><k \, | \, x'> &=&
e^{-{\displaystyle{\frac{(x-x')^2}{2L^2}}}} \ . \hspace{5.4cm} (10)\\
\int dx <k \, | \, x> <x \, | \, k'> &=&
e^{-{\displaystyle{\frac{(k-k')^2L^2}{2}}}} \ . \hspace{4.9cm} (11)
\end{eqnarray*}
Notice that the sharp delta function does not appear.

Let us find the Green's function for the scalar field
$$
( \Box \, ~+~ \, m^2 ) \, \phi(x) \, | \, x > ~=~ J(x) \, | \, x > \ .
\eqno(12)
$$
By going to $| \,k>$ basis and back to $| \, x>$ basis one can check
that
$$
\phi (x) \, | \, x> ~=~ \int dk \, dx' \, J(x') \, | \, x'>
\displaystyle{\frac{<x' \, | \, k><k \, | \, x>}{k^2 ~+~ m^2}} \ .
\eqno(13)
$$
So the usual Green's function $\int dk \, \displaystyle{\frac{e^{ik(x-x'
)}}{k^2~+~ m^2}} $ is replaced by
\newline $\int dk \, \displaystyle{\frac{<x'| k><k | x>}{k^2~+~ m^2}}$ .
This has much better properties as can be seen from the identities (10)
and (11) given earlier.  Also note that $\int dk \, | \, k><k \, |$ is
tracing over discrete states  $\sum \, |n>\,<n|$
of the interval. Hence one has introduced some type of lattice like
structure for spacetime. The sharp delta functions which are the origin
of the divergence problem are now softened.

It is now possible to write the path integral by replacing $\phi(x)$ by
$\phi(x) \, | \, x>$.
Thus formally one has (for a very simple case)
$$\int {\cal D} \phi(x) | x> exp (-i)( \int dx \, <x \, | \,
\phi(x) \, \Box \,\phi (x) |x>~~~~~~~~~~~~~~$$
\begin{eqnarray*}
&-&\int dx \, <x \, | \, J(x) \phi(x) \, | \, x>) \\
&=& exp \left( \frac{-i}{2} \int dx \, dx' \, dk \, J(x) \, | \, x>
\frac{<x \, | \, k><k \, | \, x'>}{k^2} <x' | \, J(x') \right) \ .
\hspace{.1cm} (14)
\end{eqnarray*}
Thus Feynman rules can be derived.

It is also possible to do the quantization in the canonical way. There
is no problem when it comes to introducing spinors.

To recapitulate, quantum field theory is modified by redefining the
notion of interval. This is also a modification of classical field
theory in a consistent way. The coherent state is used because of a
correspondence principle argument.
The light cone properties are improved. The new theory has far better
properties than the older theory. Further there is great formal
similarity and one can derive Feynman rules from the Path Integral
approach.

To quantize gravity is now possible. As the Green's functions have much
better properties  all the integrals occuring in the Feynman diagram
calculations will now have cutoffs (of an exponential type). Now one has
to follow Weinberg's approach and use Lorentz Invariant theory of spin
two gravitons coupling to the energy momentum tensor.

Thus the idea of Faddeev to deform the laws of physics is implemented.
The idea of Weinberg now can be used to quantize gravity. This could be
called the Weinberg  - Faddeev approach to the problems of quantum field
theory and quantum gravity.

The advantage of this approach is that experimental facts can be
accommodated. Model building is not affected. What has been made is a
general reformulation of physics and experimental details can be
incorporated in the model. This is unlike string theory which has a lot
of difficulty in agreement with experiment, which is consistent in 10 or
26 (not 4) space time dimensions and requires Kaluza - Klein theory,
Calabi Yau manifolds and several exotic ad hoc assumptions and yet makes
no contact with experimental facts enshrined in the Weinberg - Salam
Model.

It must be stressed that the approach in this paper is extremely
conservative both conceptually and mathematically.

\noindent {\bf Acknowledgements:} \ Thanks to Dr. H.S.Sharatchandra for
his encouragement and warm friendship. Thanks to L.Kannan of PPST for
his help in analysing some of the problems. Thanks to Dr.Shantha Kamath
and Mr.M.S.Kamath for giving the computer on which this paper was typed.
The work is funded by CSIR, India.

\newpage
\noindent {\bf References:}
\begin{description}
\item{[1]} {\it On the Relationship between Mathematics and Physics},
L.D.Faddeev, Asia Pacific Physics News Vol.3, June-July, 1988 pp.21-22.

\item{[2]} {\it Photons and Gravitations in S-Matrix theory:
Derivation of Charge Conservation and Equality of Gravitational and
Inertial Mass}, S.Weinberg, Phys. Rev. B1049-1056, Vol.135, 4B, 1964.

\item{~~~} {\it Photons and Gravitons in Perturbation Theory: Derivation
of Maxwell's and Einstein's Equations}, S.Weinberg, Phys. Rev.
B988-1002, Vol.138, 4B, 1965.
\end{description}
\end{document}